\documentstyle[11pt]{article}
\def\half{\mbox{$\frac{1}{2}$}}     

\begin{document}
\special{vc:  ssvcid ycn.tex 1.1 Thu 95Apr06  19:21 \quad
              ssvcid TeX'd \today \quad
              vcidtag
}
\makeatletter
\@addtoreset{equation}{section}
\makeatother
\renewcommand{\theequation}{\thesection.\arabic{equation}}


%
\title{Yilmaz Cancels Newton%
\thanks{UMD PP-95-115; gr-qc/9504050}
      }
\author{   \sc
            Charles W. Misner\\
           \em
            Department of Physics, University of Maryland
           \\ \em
            College Park MD 20742-4111 USA\\
           \rm
         e-mail: \tt misner@umail.umd.edu
        }
\date{28 April 1995}
\maketitle

\begin{abstract}
    A central tenet of the new theory of gravity proposed by H.
Yilmaz is the inclusion of a gravitational stress-energy tensor
$-{t_\mu}^\nu$ along with the matter stress-energy tensor
${T_\mu}^\nu$ on the right hand side of the Einstein field equations.
This change does not effect the Newtonian limit of the field equations since
these terms are quadratic in potential gradients.  From the Bianchi
identities, however, important changes appear in any equations of
motion consistent with these field equations.  For matter described
as a perfect fluid, and with Yilmaz's choice of signs when introducing
these quadratic terms, we find that the Euler hydrodynamic equation
in the Newtonian limit is modified to remove all gravitational
forces.  This allows, e.g., a solar system in which the Sun and the
planets are permanently at rest, but does not explain how fluid bodies
such as the Sun or Jupiter could form or be prevented from dispersing.
\hfill\\[1.5ex]
\end{abstract}

\section{Introduction}
    Recent publications by Yilmaz \cite{Yil92,Yil94} claim that
General Relativity (GR) is inconsistent and that its Newtonian limit
is unsatisfactory. Because these claims have resulted in some widely
distributed notice \cite{SciNews,Science}, and because they may be
having an invalid influence on experimental proposals \cite{All94},
it seems appropriate to restate that they are incorrect.  But the
straightforward and long established arguments (or more rigorous
recent work \cite{FrR94}) showing GR to have a good Newtonian limit
leave little room for including Newtonian
stresses in still another way in the field equations, as Yilmaz has
proposed.  We will see here that the Yilmaz theory makes changes to
GR that are so profound they remove all gravitational
interactions at the Newtonian level in any exact solution of the
field equations of the modified theory.

    In notation that differs in detail from that used by Yilmaz in
his papers, but adopts the MTW conventions~\cite{MTW73}, the Einstein
field equations are
    \begin{equation}\label{e-Ein}
        R^{\mu\nu} - \half g^{\mu\nu} R
          \equiv  G^{\mu\nu} = (8 \pi G / c^4) T^{\mu\nu}
        \quad ,
    \end{equation}
    where all components of $T^{\mu\nu}$ have units of energy density
(i.e., pressure) and this tensor describes the stress-energy of all
non-gravitational fields.  In the Yilmaz theory this equation is
modified to read
    \begin{equation}\label{e-Yil}
        G^{\mu\nu} = (8 \pi G / c^4) (T^{\mu\nu} - t^{\mu\nu})
        \quad ,
    \end{equation}
    where in every instance $t^{\mu\nu}$ as used here differs in sign from
the quantity Yilmaz represents by this symbol. The exact definition of
the Yilmaz gravitational stress-energy tensor
$-t^{\mu\nu}$ is in some cases unclear, but explicit expressions
have been given in situations that include the Newtonian
limit.  In suitable (harmonic) coordinate systems chosen by Yilmaz
the definition becomes ${t_\mu}^\nu = {u_\mu}^\nu$ where ${u_\mu}^\nu$
is the Einstein stress-energy pseudotensor; my meaning of
${t_\mu}^\nu$ avoids a minus sign in this definition, as well as in
the relationship~(\ref{e-YNt}) below of ${t_\mu}^\nu$ to the Newtonian
gravitational stresses.

\section{Relativistic Matter}
    To compare the GR and Yilmaz theories in the Newtonian limit one
needs a theory of matter so there is something to exhibit
gravitational interactions.  I will take for a theory of matter the
relativistic hydrodynamics of a perfect fluid.  [Newtonian
hydrodynamics with gravitational forces is summarized in
appendix~\ref{s-N}.]  This is a suitable test case as it avoids
dealing with the singularities of point particles, yet includes some
nongravitational (pressure gradient) forces which are needed in the
Newtonian theory for static solutions to exist.  In this case one
takes the stress-energy tensor of matter to be
    \begin{equation}\label{e-GRpf}
        T^{\mu\nu} = \rho c^2 u^\mu u^\nu + P (g^{\mu\nu} + u^\mu u^\nu)
        \quad .
    \end{equation}
    The fluid 4-velocity is normalized by the condition
    \begin{equation}\label{e-GRnorm}
        g_{\mu\nu} \, u^\mu u^\nu = -1
    \end{equation}
    so that in quasi-rectangular coordinates the 4-velocity can be
written
    \begin{equation}\label{e-u}
         u^\mu =  (1, {\bf v}/c ) \gamma
    \end{equation}
where $v^k = dx^k/dt$ is the coordinate velocity of the fluid
particles, $x^0 = ct$ is the fourth coordinate in tensor
expressions, and $\gamma$ will be determined by the
normalization~(\ref{e-GRnorm}).  All components  $g_{\mu\nu}$ and
$u^\mu$ are dimensionless, as $\gamma$ will then be.

    In the Yilmaz theory the hydrodynamic equations have
gravitational influences that differ from those in GR.  In this case
of a matter theory with only four independent fields ($\rho$ and the
three $v^k$ with $P(\rho)$ given by an equation of state) the
Bianchi identities which enforce local energy and momentum
conservation constrain the matter equations of motion sufficiently to
determine them completely.  The contracted Bianchi identities
${G^{\mu\nu}}_{;\nu} = 0$ from equation~(\ref{e-Yil}) imply
$(T^{\mu\nu} - t^{\mu\nu})_{;\nu} = 0$ in the Yilmaz theory.

    When split into components parallel and perpendicular to the
fluid 4-velocity $u^\mu$ the local conservations equations
$(T^{\mu\nu} - t^{\mu\nu})_{;\nu} = 0$ are an energy equation
    \begin{equation}\label{e-Rcontin}
        (\rho c^2 u^\mu + u_\nu t^{\nu\mu})_{;\mu} =
                -P {u^\nu}_{;\nu} + t^{\mu\nu} u_{\mu ; \nu}
        \quad ,
    \end{equation}
    from $-u_\mu(T^{\mu\nu} - t^{\mu\nu})_{;\nu} = 0$
and the generalized Euler equation
    \begin{equation}\label{e-REuler}
        (\rho c^2 + P){u^\mu}_{;\nu} u^\nu =
                    -(g^{\mu\nu} + u^\mu u^\nu) (P_{;\nu}
                             -{{t_\nu}^\alpha}_{;\alpha})
    \end{equation}
    from $(\delta^\mu_\nu + u^\mu u_\nu)
(T^{\nu\alpha} - t^{\nu\alpha})_{;\alpha} = 0$.
    The terms involving $t^{\mu\nu}$ are absent for the Einstein
theory of gravity. In equation~(\ref{e-REuler}) this allows geodesic
motion ${u^\mu}_{;\nu} u^\nu = 0$ in GR when the pressure gradient
forces on the right are also absent.  Although the $t^{\mu\nu}$
terms, like the pressure terms,
should be negligible in the Newtonian limit in the gravitational
field equations~(\ref{e-Yil}) and also negligible in the equation of
continuity (\ref{e-Rcontin}) they may, like the pressure gradient
terms, be important in the Newtonian limit of the Euler
equation~(\ref{e-REuler}).  We will find that they cancel the terms
from $\rho c^2{u^\mu}_{;\nu} u^\nu$ that in Einstein's theory give
the Newtonian gravitational force term in the Euler equation.

    The generalized Euler equation~(\ref{e-REuler}) is not optional
or modifiable in the Yilmaz theory.  If a $t^{\mu\nu}$ term is
inserted in the Einstein equations, the Bianchi identities show that
it must appear in the hydrodynamic equations of motion satisfied in
any exact solution.  Thus to leading order it must be satisfied by
any motion that approximates an exact solution of
equations~(\ref{e-Yil}) with a hydrodynamic $T^{\mu\nu}$
stress-energy tensor.  This argument makes
no use of any additional equations that may be part of the Yilmaz
theory, as we have to this point made use only of the field
equations~(\ref{e-Yil}) and the assumption that the theory is
applicable to selfgravitating systems of fluid objects with the
stress-energy tensor~(\ref{e-GRpf}).

\section{An Exact Solution}
    Yilmaz has given an exact solution of his field equations
\cite{Yil94,All94} which
I derive here to verify that, in spite of changes in notation and
sign conventions, I am stating equations from his theory correctly.
When the consequences~(\ref{e-REuler}) of the field equations are
considered in this example one sees a failure to reproduce Newtonian
gravitational effects.
The metric Yilmaz proposes is
    \begin{equation}\label{e-GRN}
        ds^2 = -e^{2 \Phi} d(ct)^2 + e^{-2 \Phi} (dx^2 + dy^2 + dz^2)
         \quad .
    \end{equation}
For this choice of metric and coordinates Yilmaz evaluates his
gravitational stress-energy tensor $-{t_\mu}^\nu$ to give
    \begin{equation}\label{e-Yt}
        {t_\mu}^\nu = \lambda (c^4/4 \pi G)[\Phi_{;\mu} \Phi^{;\nu}
               - \half \delta_\mu^\nu \Phi_{;\alpha}\Phi^{;\alpha}]
    \end{equation}
with $\lambda = 1$.  We have inserted the tag $\lambda$ here so that
by setting $\lambda = 0$ in later equations we can recover the
corresponding equation in Einstein's theory.

    I now use the computer algebra aid {\em GRTensor II\/}~\cite{MPK94,MaV94}
to construct the Yilmaz field equations in this case.  The result is
    \begin{eqnarray}
       Y^{00}:&  & 2\Phi_{,kk} + (\lambda-1)\Phi_{,k}\Phi_{,k}
                + (\lambda + 3)e^{-4\Phi}(\Phi_{,0})^2 \nonumber\\
          & = & (8 \pi G/c^2) [\rho \gamma^2
                + (P/c^2)(e^{-2\Phi} - \gamma^2)] \quad ,
                \label{e-Y00}\\[1ex]
       Y^{0j}:&  &  -2\Phi_{,j0}
                -   2  (\lambda-1)\Phi_{,j}\Phi_{,0}\nonumber\\
          & = & (8 \pi G/c^3) \gamma^2 v^j (\rho + P/c^2) \quad ,
          \label{e-Y0j}\\[1ex]
       Y^{ij}:&  & (1 - \lambda) e^{4\Phi}
                [\delta^{ij}\Phi_{,k}\Phi_{,k} - 2\Phi_{,i}\Phi_{,j}]
                \nonumber\\
          &   &\quad  + \delta^{ij}
            [2\Phi_{,00} + (\lambda-5)(\Phi_{,0})^2]\nonumber\\
          & = & (8 \pi G/c^4) [\gamma^2 v^i v^j (\rho + P/c^2)
             +  \delta^{ij} P e^{2\Phi}] \quad .
             \label{e-Yij}
    \end{eqnarray}
    The left hand sides of these equations are the gravitational
quantities $Y^{\mu\nu} \equiv G^{\mu\nu} + (8 \pi G/c^4) t^{\mu\nu}$
while all the matter terms are collected on the right.

    To find a static solution we set $\lambda =  1$ (Yilmaz theory),
drop all time derivatives, set   $v^j = 0$, and from the
normalization condition~(\ref{e-GRnorm}) get $\gamma^2 = e^{-2\Phi}$.
The resulting equations are
    \begin{eqnarray}
       Y^{00}:\quad\: 2\Phi_{,kk}
          & = & (8 \pi G/c^2) \rho e^{-2\Phi} \quad ,
          \label{e-Yx00}\\
       Y^{0j}:\qquad\quad 0
          & = & 0 \quad ,\\
       Y^{ij}:\qquad\quad 0
          & = & (8 \pi G/c^4) \delta^{ij} P e^{2\Phi} \quad .
          \label{e-Yxij}
    \end{eqnarray}
    To construct solutions to these equations one chooses any desired
density distribution $\rho_{\rm eff}(x,y,z)$ representing one or more
fluid bodies and then solves the Poisson equation $U_{,kk} = 4 \pi G
\rho_{\rm eff}$. Then the choice $\Phi = U/c^2$ reduces
equation~(\ref{e-Yx00}) to the form $\rho_{\rm eff} = \rho
e^{-2U/c^2}$ which serves to display the invariant density $\rho=
u_\mu u_\nu T^{\mu\nu}/c^2$ in terms of the given effective density.
To solve equation~(\ref{e-Yxij}) one must set $P=0$.

    The existence of this exact solution is disastrous for the
Yilmaz theory.  Any equation of motion for fluid matter adjoined to
the Yilmaz field equations~(\ref{e-Yil}) cannot differ from the
generalized Euler equation~(\ref{e-REuler}) since that equation is
just a combination of the field equations and their derivatives. Thus
this static solution with $P=0$ satisfies also the fluid equations of
motion~(\ref{e-REuler}) in any consistent theory incorporating the
Yilmaz field equations~(\ref{e-Yil}).  One such solution could
represent two fluid planets (Jupiter and Saturn) placed near a fluid
Sun with none of these bodies in motion.  No gravitational forces
act to accelerate any of these bodies, no mechanical (pressure) forces
are needed to balance gravity, and
the only gravitational effect is the renormalization between
effective density $\rho_{\rm eff}$ and local invariant density
$\rho$.

    The Einstein equations under these same static assumptions have no exact
fluid solution in agreement with the Newtonian impossibility of a
motionless solar system.  The usual Oppenheimer-Volkoff static fluid solutions
(spherical symmetry only) are obtained by allowing $-g_{00}$ and
$g^{xx}$ to differ and do require pressure gradients to support the matter
against its own gravitational fields.  See, for example, \cite[\S
23.7]{MTW73}.  To see that allowing additional independent components
in $g_{\mu\nu}$ does not help in the Yilmaz theory we treat the
Newtonian limit below without assuming static configurations
nor restrictions beyond harmonic coordinate conditions on the metric
components.

\section{Newtonian limit}
    In the Newtonian limit one assumes that velocities are small,
$|v/c| \ll 1$, that gravitational potentials $g_{\mu\nu}$ are near
their Minkowski values, and that pressures or other mechanical
stresses are negligible compared to energy densities $|P| \ll \rho
c^2$.  By arguments known since Einstein and Hilbert, one finds that
only one combination of the $g_{\mu\nu}$ is generated by the dominant
term in $T^{\mu\nu}$ which is $\rho c^2 u^0 u^0$.  (See, e.g.,
\cite[\S 18.4]{MTW73}.)  Neglecting terms in $g_{\mu\nu}$ smaller
than this leading term by factors of $v/c$ or $\Phi$ one finds that
the metric is the same as equation~(\ref{e-GRN}) in which one may set
$e^{\pm 2\Phi} = 1 \pm 2 \Phi$.  This metric will be time dependent
with $\Phi_{,0} = \partial\Phi/\partial ct$ smaller than $\Phi_{,k}$
by a factor of $v/c$ where $v$ is the velocity of the matter
sources.  [Consider, for example, the Newtonian potential $U = -GM/|{\bf
r - v}t|$ of a moving point mass.]   The equality of $-g_{00}$ and
$g^{xx}$ is deduced rather than assumed, and holds only as a first
approximation.  These arguments hold equally well
for the Yilmaz theory or for the Einstein theory since $t^{\mu\nu}$
is of order $(\partial \Phi)^2$ and thus smaller than the leading
$\partial^2 \Phi$ terms since $|\Phi| \ll 1$.  Thus both the Yilmaz
theory and the Einstein theory lead to
    \begin{equation}\label{e-NR}
       2\Phi_{,kk}  =  (8 \pi G/c^2) \rho
    \end{equation}
    from the $G^{00}$ equation, while all terms in the other field
equations are negligible in comparison.  The solution of this
equation is
    \begin{equation}\label{e-Nlim}
       \Phi  =  U/c^2
    \end{equation}
    where $U$ is the Newtonian potential satisfying
equation~(\ref{e-Poisson}) for the same matter distribution
$\rho(x,y,z,t)$.

    Since both the Einstein and Yilmaz field equations give the same
metric in the Newtonian limit, it had long been assumed that the
Yilmaz theory had a satisfactory Newtonian limit.  But Yilmaz
\cite{Yil94} has recently suggested that the theories differ in this limit,
prompting this study which localizes the difference to the equations
of motion required by the Bianchi identities.  Thus we proceed to
study approximations to equations~(\ref{e-Rcontin}) and
(\ref{e-REuler}).

    The reduction of the relativistic continuity
equation~(\ref{e-Rcontin}) is relatively straightforward using the
metric~(\ref{e-GRN}).  The terms on the right by which restmass is
changed by mechanical or gravitational work are unimportant for
Newtonian strength forces, and in the mass current vector $\rho c^2
u^\mu + u_\nu t^{\nu\mu}$ the gravitational binding energy contribution
to the rest mass from $t^{\nu\mu}$ is also small.  This leaves
$(\rho c u^\mu)_{;\mu} = 0$ in which the approximation $u^\mu = (1,
v^j/c)$ will suffice.  All terms involving $\partial \Phi$ are
smaller by a factor $\Phi$ than those in the Newtonian equation and
thus negligible, so equation~(\ref{e-Ncontin}) results.

    The relativistic Euler equation~(\ref{e-REuler}) similarly
simplifies.  (The $\mu = 0$ equation is negligible.) The left hand
side with $u^0 = 1$ and $u^i = v^i/c$ reads $\rho c^2 {u^i}_{;\nu}
u^\nu = \rho [(\partial v^i/\partial t) + v^k {v^i}_{,k} +
{\Gamma^i}_{00}c^2]$ and the right hand side is $-P_{,i} +
{t^{ik}}_{,k}$.  But when one makes the identification $\Phi = U/c^2$
from the field equations the Yilmaz term from equation~(\ref{e-Yt})
is just given by the Newtonian stresses~(\ref{e-Ngstress})
    \begin{equation}\label{e-YNt}
        t^{ik} = \lambda t_{\rm N}^{ik}
    \end{equation}
    so from equation~(\ref{e-Ngforce}) one has ${t^{ik}}_{,k} =
\lambda \rho U_{,i}$.  Also one can evaluate ${\Gamma^i}_{00}c^2 =
U_{,i}$ from the metric and $\Phi = U/c^2$.  The result then is
    \begin{equation}\label{e-YNEuler}
       \rho [\partial v^i/\partial t + {v^i}_{,k} v^k] =
                       - P_{,i}  - \rho(1-\lambda) U_{,i} \quad .
    \end{equation}
    This is the desired Newtonian result (\ref{e-NEuler}) in the
Einstein case $\lambda = 0$ but lacks any Newtonian gravitational
force in the Yilmaz case $\lambda = 1$.
\vspace{2\abovedisplayskip}

\noindent
{\Large\sc Appendix}
\appendix
\vspace{-\abovedisplayskip}
\section{Newtonian Benchmark}\label{s-N}
    The Newtonian theory we expect to recover as a limiting case of
any relativistic theory will be taken to be the hydrodynamics of a
perfect fluid.  This theory has as its gravitational field equation
the Poisson equation
    \begin{equation}\label{e-Poisson}
         U_{,kk} = 4 \pi G \rho
    \end{equation}
    for the generation of gravitational potentials, the continuity
equation
    \begin{equation}\label{e-Ncontin}
       \partial \rho/\partial t + (\rho  v^k)_{,k} = 0
    \end{equation}
    which shows how fluid motions $v^k$ affect the density, and the
Euler equation
    \begin{equation}\label{e-NEuler}
        dv^i/dt \equiv \partial v^i/\partial t + {v^i}_{,k} v^k =
                                - U_{,i} - (1/\rho)P_{,i}
    \end{equation}
    which shows how the force per unit mass changes the momentum per
unit mass $v^i$.  (Note that for a point mass one has $U = -GM/r$
which is the opposite sign from that used for the symbol $U$ in \cite[\S
39.7]{MTW73}.)  To have a complete deterministic theory these equations must
be supplemented by an equation of state $\rho = \rho (P)$.  Thermal
properties including viscosity and heat transfer are neglected here,
and the equation of state is assumed not to involve the temperature.

    The Euler equation can be replaced in the Newtonian theory by an
equivalent form in which all forces, not just pressure forces, are
described by a stress tensor or momentum flux tensor
    \begin{equation}\label{e-Nstress}
        T_{\rm N}^{ij} = \rho v^i v^j + P^{ij} + t_{\rm N}^{ij}
    \end{equation}
    where $\rho v^i v^j$ give the convective transport of momentum and
$P^{ij}$ are the mechanical stresses.  (When two regions are in
contact the forces they exert on each other represent, by action
equals reaction, a momentum gain for one region and a corresponding loss for
the other.  Thus a contact force per unit area is a momentum transfer
rate per unit contact area.)  The gravitational stresses
    \begin{equation}\label{e-Ngstress}
        t_{\rm N}^{ij} = (1/4 \pi G)[U_{,i} U_{,j} -
                \half \delta^{ij} U_{,k} U_{,k} ]
    \end{equation}
    are chosen so that, using the Poisson equation, they give the
required gravitational force per unit volume:
    \begin{equation}\label{e-Ngforce}
       - {t_{\rm N}^{ik}}_{,k} = -(1/4 \pi G) U_{,i} U_{,kk} = -\rho U_{,i}
            \quad .
    \end{equation}
    For application to fluids one sets $P^{ij} = P \delta^{ij}$ and
finds that $-{P^{ik}}_{,k} = -P_{,i}$ is the pressure gradient force
per unit volume.  Conservation of momentum is then expressed by
    \begin{equation}\label{e-Nmom}
        \partial (\rho v^i)/\partial t +{T_{\rm N}^{ik}}_{,k} = 0
    \end{equation}
    which treats momentum on a ``per unit volume'' basis in contrast
to the previous Euler equation~(\ref{e-NEuler}) which is on a ``per
unit mass'' basis. It is straightforward, using the equation of
continuity, to reduce equation~(\ref{e-Nmom}) to
equation~(\ref{e-NEuler}).

\end{document}